\documentclass[notitlepage,english,aps,floats,onecolumn,showpacs,nofootinbib,floatfix]{revtex4-1}
\usepackage{pslatex}
\usepackage[T1]{fontenc}
\usepackage[latin1]{inputenc}
\usepackage{graphicx}
\usepackage{epsfig}
\usepackage{longtable}
\usepackage{float}
\usepackage{calc}
\usepackage{ifthen}
\usepackage{amsmath}
\usepackage{hyperref}
\usepackage{amssymb}

\usepackage{color}

{
{
{
\newcommand{\bea}{\begin{eqnarray}}
\newcommand{\eea}{\end{eqnarray}}

\newcommand{\nc}{\newcommand}
\nc{\renc}{\renewcommand}
\nc{\eqs}[2]{\mbox{Eqs.~(\ref{#1},\,\ref{#2})}}
\nc{\eq}[1]{\mbox{Eq.~(\ref{#1})}}
\nc{\figs}[2]{\mbox{Figs.~(\ref{#1},\,\ref{#2})}}
\nc{\fig}[1]{\mbox{Fig~.(\ref{#1})}}
\nc{\be}[1]{\begin{equation} \mbox{$\label{#1}$}}
\nc{\ee}{\vspace{0.1cm}\end{equation}}

\newcommand{\bean}{\begin{eqnarray*}}
\newcommand{\eean}{\end{eqnarray*}}

\def\GeV{{\rm \ GeV}}

\def\TeV{{\rm \ TeV}}

\def\lae{\;^{<}_{\sim} \;} \def\gae{\; ^{>}_{\sim} \;}

\begin{document}

\title{Vector-like Quark Stabilised Higgs Inflation: Implications for Particle Phenomenology, Primordial Gravitational Waves and the Hubble Tension} 

\author{John McDonald }
\email{j.mcdonald@lancaster.ac.uk}
\affiliation{Dept. of Physics,  
Lancaster University, Lancaster LA1 4YB, UK}

\begin{abstract}  The Standard Model (SM) Higgs potential is likely to be metastable, in which case Higgs Inflation requires an extension of the SM to sufficiently stabilise the Higgs potential. Here we consider stabilisation by adding $n_{Q} \leq 3$ Vector-Like Quarks (VLQs) of mass $m_{Q}$. We consider isosinglet $T$ vector quarks transforming under the SM gauge group as $({\bf 3}, {\bf 1}, 2/3)$ and $B$ vector quarks transforming as $({\bf 3}, {\bf 1}, -1/3)$. Requiring stability of the finite temperature effective potential after instant reheating, and assuming that the $t$-quark mass $m_{t}$ equals the mean value of its experimental range,  we find that the upper bounds on 
$m_{Q}$ for $T$ quarks are 5.8 TeV (for $n_{Q} = 2$) and 55 TeV (for $n_{Q} = 3$). The corresponding absolute stability upper bounds are 4.4 TeV and 29 TeV. For $n_{Q} = 1$ there is stability only for $m_{t}$ at its -2-$\sigma$ value, in which case $m_{Q} \leq 1.6 \TeV$ for one $T$ quark. The upper bounds are generally smaller for $B$ vector quarks, with finite temperature stability  for $m_{Q}$ less than 2.8 TeV  (for $n_{Q} = 2$), 18 TeV (for $n_{Q} = 3$) and 1.0 TeV (for $n_{Q} = 1$). The upper bounds on $m_{Q}$ are sensitive to the $t$-quark mass, becoming smaller as $m_{t}$ increases. The inflation predictions depend upon the conformal frame in which the model is renormalised. For renormalisation in the Einstein frame (Prescription I) the predictions are almost indistinguishable from the classical values: $n_s = 0.966$ and $r = 3.3 \times 10^{-3}$. In this case  the stability upper bounds on $m_{Q}$ apply. Renormalisation in the Jordan frame  (Prescription II) predicts larger values of $n_{s}$ and $r$, with $n_{s}$ generally in the range 0.980 to 0.990 and $r$ of the order of 0.01. The predicted range of $n_{s}$ is consistent with the CMB range obtained in Hubble tension solutions which modify the sound horizon at decoupling, whilst the predicted values of $r$ will be easily observable by forthcoming CMB experiments. The observational upper bound on $r$ generally imposes a stronger constraint on $m_{Q}$ in Prescription II than the requirement of stability, with the $T$ quark upper bound equal to 2.4 TeV for $n_{Q} = 2$ and 13 TeV for $n_{Q} = 3$, assuming $m_{t}$ equals its mean value. $n_{Q} = 1$ is generally ruled out by the large value of $r$. The $m_{Q}$ upper bounds rapidly decrease with decreasing $r$. We conclude that VLQ-stabilised Higgs Inflation with Prescription II renormalisation favours 1-10 TeV vector-like quarks that will be accessible to future colliders, and predicts a tensor-to-scalar ratio that will be observable in forthcoming CMB experiments and values of $n_{s}$ that favour an early-time solution to the Hubble tension.

\end{abstract} 
 \pacs{}
 
\maketitle

\section{Introduction} 

The electroweak vacuum is likely to be metastable due to quantum corrections \cite{unst1,butt,unst3,sher,frog,stein1,litimvs} \footnote{In \cite{litimvs} it was found that, using 2023 PDG inputs, there is only a small region of the 4-$\sigma$ ellipse in $m_{t}$ and $\alpha_{s}(M_{z})$ for which stability is possible, and using correlated CMS inputs, which have larger errors but take into account correlations between the measurements, there is only a very small region of the 2-$\sigma$ ellipse for which stability is  possible. In \cite{stein1} it was also found that only a small region of the 2-$\sigma$ ellipse is still compatible with stability. A factor of 2 improvement in the errors will exclude stability to 5-$\sigma$ \cite{litimvs}.}.  In the case of the SM this does not present any cosmological or phenomenological problem, as the Universe can naturally evolve into the electroweak vacuum once finite temperature evolution is taken into account \cite{fte,fte2}. However, in the case of Higgs Inflation \cite{bs,salopek}, it is essential that the Higgs potential is sufficiently stable to support inflation, which requires that the instability scale is greater than the Higgs field during inflation, of the order of $M_{Pl}/\sqrt{\xi}$, where $\xi$ is the non-minimal 
coupling. To achieve this, it is likely that additional particles must be added to the SM \footnote{It is possible to have Higgs Inflation in the unmodified SM if the SM instability scale is larger than the Higgs field during inflation, which requires a low value for $m_{t}$. In \cite{mass} (see also \cite{wen2}) it is proposed that this is possible if the PDG pole mass from the $t \overline{t}$ production cross-section is considered rather than the direct measurement value used here. While the direct measurement value has smaller errors, there is a theoretical uncertainty between the Monte Carlo generator $t$-quark mass, which is extracted from direct measurements of the kinematics of $t \overline{t}$ events, and the pole mass. This is due to non-perturbative effects that are difficult to quantify \cite{pdg1,nason,hoang}. For a discussion of the $t$-quark mass, uncertainties in its determination and its implications for electroweak vacuum stability, see \cite{tqu}. } 
 Here we consider the addition of vector-like fermions, focusing on isosinglet vector-like quarks (VLQs) in the $({\bf 3}, {\bf 1}, Y_{Q})$ representation. Vector-like fermions are anomaly-free and allow a mass term. We will specifically focus on vector-like quarks in the same repesentation as SM quarks: $T$ vector-like quarks transforming as $({\bf 3}, {\bf 1}, 2/3)$ and $B$ vector-like quarks transforming as $({\bf 3}, {\bf 1}, -1/3)$.  These can mix with the SM quarks and so decay, as is necessary to avoid cosmologically excluded stable coloured 
particles \cite{cosmo}.  It is known that vector-like fermions can stabilise the Higgs potential through modification of the renormalisation group (RG) running of the gauge couplings \cite{vlqs1,vlqs2,litim, litim2}. TeV-scale fermions have also been motivated by vacuum selection considerations in \cite{stein1} and \cite{stein2}.   

In this paper we will determine the upper bounds on the VLQ mass $m_{Q}$ from the requirement of sufficient Higgs stability for inflation and the predictions of the model for inflation observables. An important distinction should be made between absolute stability of the zero temperature Higgs potential and stability of the finite temperature effective potential. For successful Higgs Inflation, it is only necessary that the Universe evolve into the electroweak vacuum following inflation and reheating. We will show that the upper bounds on $m_{Q}$ from the finite temperature effective potential are weaker than those from absolute stability of the Higgs potential.  

In the following we will consider $n_{Q}$ VLQs of type $T$ or $B$. We will run the 3-loop SM RG equations \cite{butt}, modified to include the Higgs propagator suppression due to the non-minimal coupling \cite{wilczek, star, corr1}, together with the leading order 2-loop corrections due to the VLQs \cite{vlqs1}.   

The upper bounds on $m_{Q}$ from stability are independent of the renormalisation frame of the quantum corrections to the Higgs potential. However, the conformal frame in which the model is renormalised is important when calculating the inflation observables: the scalar spectral index $n_{s}$ and tensor-to-scalar ratio $r$. Two cases are commonly considered \cite{presc1,presc2}; renormalisation in the Einstein frame (Prescription I) and renormalisation in the Jordan frame (Prescription II). These frames correspond to different UV completions of the theory and should therefore be considered as different Higgs Inflation models \cite{uvframe}. We will show that whereas  Prescription I predicts essentially the same values as classical Higgs Inflation, Prescription II predicts values of $n_{s}$ and $r$ that are considerably larger than classical Higgs Inflation.  This allows Higgs Inflation to be compatible with the CMB spectral index of Early Dark Energy (EDE) solutions of the Hubble tension, which typically require larger values of $n_{s}$ to fit the observed CMB than $\Lambda$CDM \cite{taka,giare,ede}. In addition, the predicted primordial gravitational waves are close to the present observational upper bound and will be easily observable by forthcoming CMB experiments \cite{lite}.

The paper is organised as follows. In Section II we review the Higgs Inflation model and VLQ stabilisation. In Section III discuss the quantum corrections to the Higgs potential. In Section IV we discuss the finite temperature effective potential. 
In Section V we present our results and in Section VI we discuss our conclusions.

\section{Higgs Inflation with Vector-Like Quarks} 
 
We first review the essential aspects of Higgs Inflation and its modification via VLQs. The action of the model in the Jordan frame is 
\be{e1}      S = \int d^{4} x \sqrt{-g} \left[ \left(M_{Pl}^{2} + \xi \phi^{2} \right) \frac{R}{2}   - \frac{1}{2} \partial_{\mu} \phi \partial^{\mu} \phi  - V(\phi) + \hat{ {\cal L} } \right]  ~,\ee
where $\hat{{\cal L}} $ is the Lagrangian of the SM and VLQ fields excluding the Higgs kinetic and potential terms. To analyse inflation and the post-inflation era, we transform the action to the Einstein frame via a conformal transformation $\tilde{g}_{\mu \nu} = \Omega^{2} g_{\mu\nu}$, where 
\be{e2} \Omega^{2} = \left(1 + \frac{\xi \phi^{2}}{M_{Pl}^{2}} \right)   ~.\ee 
The Einstein frame action is then 
\be{e4}  S = \int d^{4} x \sqrt{-\tilde{g}} \left[ \frac{M_{Pl}^{2}}{2} \tilde{R} -\frac{1}{2 \Omega^{2}} \left(1 + \frac{6 \xi^{2} \phi^{2} }{\Omega^{2} M_{Pl}^{2}} \right) \partial_{\mu} \phi \partial^{\mu} \phi 
 - V_{E}(\phi) + \frac{ \hat{\cal{L}}}{\Omega^{4}}  \right]   ~,\ee
where $V_{E}(\phi) = V(\phi)/\Omega^{4}$ is the Higgs potential in the Einstein frame.

At field values $\phi >  \phi_{c} = M_{Pl}/\sqrt{\xi} \sim 10^{16} \GeV$, the conformal factor $\Omega$ strongly deviates from 1 and the classical Einstein frame potential has a plateau suitable for inflation. To obtain the predictions for the model, we will
numerically solve for the end of inflation, $\sigma_{end}$, defined as when 
either $|\eta(\sigma)| > 1$ or $\epsilon(\sigma) > 1$ first occurs as $\sigma$ decreases. Here $\sigma$ is the canonically normalised inflaton, which is related to $\phi$ by
\be{e13a} \frac{d \sigma}{d \phi} = \frac{1}{\Omega} \left(1 + \frac{6 \xi^{2} \phi^{2} }{\Omega^{2} M_{Pl}^{2}} \right)^{1/2}  ~.\ee
We then numerically integrate for the number of e-foldings of inflation as a function of $\sigma$ 
\be{e14} N = - \frac{1}{M_{Pl}^{2}} \int_{\sigma}^{\sigma_{end}}\frac{V_{E}(\sigma)}{V'_{E}(\sigma)} d \sigma   ~\ee 
to determine $\sigma$  and so $\phi$ at the pivot scale $N_{*}$. 
The scalar spectral index, tensor-to-scalar ratio and curvature perturbation power spectrum are calculated in the standard way 
\be{e15} n_{s} = 1 + 2 \eta - 6 \epsilon  ~,\ee 
\be{e16} r = 16 \epsilon ~,\ee 
and
\be{e17} {\cal P}_{\zeta} =  \frac{V_{E}(\sigma)}{24 \pi^{2} \epsilon M_{Pl}^{4}} ~,\ee 
where the slow-roll parameters are $\eta= M_{Pl}^{2} V_{E}''(\sigma)/V_{E}(\sigma)$ and $\epsilon = (M_{Pl}^{2}/2)(V_{E}'(\sigma)/V_{E}(\sigma))^{2}$, where primes denote derivatives with respect to $\sigma$.

Once quantum corrections are included, the SM potential becomes negative once $\phi \gae \Lambda$ and Higgs Inflation is no longer possible if $\Lambda < \phi_{c}$. In the next section we will show that this is true over the whole 2-$\sigma$ range of $m_{t}$. The introduction of VLQs modifies the RG evolution primarily by modifying the running of the strong gauge coupling $g_{3}$, which in turn modifies the running of the $t$-quark Yukawa $y_{t}$ and so the running of $\lambda_{h}$. This can be seen by considering the contribution to the 1-loop RG equations due purely to $y_{t}$ and $g_{3}$ \cite{unst3,vlqs1} 
\be{rg1} \mu \frac{\partial \lambda_{h}}{\partial \mu} = - 6 y_{t}^{4} + ...    ~\ee   
\be{rg2} \mu \frac{\partial y_{t}}{\partial \mu} = - 8 y_{t} g_{3}^{3} + \frac{9}{2} y_{t}^{3} + ...    ~\ee   
\be{rg3} \mu \frac{\partial g_{3}}{\partial \mu} = - 7 g_{3}^{3} + \frac{2}{3} n_{Q} g_{3}^{3} + ...    ~\ee   
Increasing $n_{Q}$ reduces the rate of decrease of $g_{3}$ with $\mu$ and so increases $g_{3}$ at a given $\mu$, which reduces the rate of increase of $y_{t}$ and so reduces $y_{t}$ for a given $\mu$. This in turn reduces the rate of decrease of $\lambda_{h}$ with $\mu$, decreasing the instability of the Higgs potential.

\section{Quantum corrections to the Higgs potential}

\subsection{The Metastable Standard Model Higgs potential}

To demonstrate the instability of the SM Higgs potential, we first compute the SM Higgs potential using the RG equations and initial $\overline{{\rm MS}}$ values of the SM couplings that we will later use for VLQ-stabilised Higgs Inflation. We run the 3-loop SM RG equations given in \cite{butt} and use the relation between the $t$-quark mass and the $\overline{{\rm MS}}$ $t$-quark coupling at $\mu = m_{t}$ \cite{butt}, 
\be{r1} y_{t} = 0.93690 + 0.00556 \left(m_{t} - 173.34 \GeV\right)  ~.\ee 
For the range of t-quark mass, we use the 2022 PDG release direct measurement value, $m_{t} = 172.69 \pm 0.30 \GeV$ \cite{pdg}. Since metastability is less sensitive to the errors in the other SM inputs (although there is a significant dependence on the strong gauge coupling), in this analysis we will use the mean values for those quantities given in \cite{butt}:  
$g_{3} = 1.1666$, $g = 0.64779$, $g' = 0.35830$ and $\lambda_{h} = 0.12604$. 

In Figure 1 we show the Higgs self-coupling $\lambda_{h}(\mu)$, calculated at the RG scale $\mu = \phi$, for the mean, -1-$\sigma$ and -2-$\sigma$ values of $m_{t}$, corresponding to progressively weaker instability. In all cases the potential runs to negative values. 

In Figure 2 we show the Higgs potential calculated using the 1-loop Coleman-Weinberg (CW) correction at $\mu = \phi$. In the figure we show ${\rm log_{10}} |V(\phi)|$ multiplied by the sign of $V(\phi)$, which is useful for visualising the negative gap in the Higgs potential. For all $m_{t}$ within the 2-$\sigma$ range the potential is metastable, with the instability scale given by $\Lambda= 2.2 \times 10^{11}$ GeV, $7.9 \times 10^{11}$ GeV and $3.0 \times 10^{12}$ GeV for the mean, -1-$\sigma$ and -2-$\sigma$ values of $m_{t}$, respectively.  The potential eventually becomes positive again, at $\phi > \phi_{upper} = 10^{29} \GeV$ for the mean value of $m_{t}$, where $\phi_{upper}$ is the upper bound of the negative gap in the potential. Therefore it is likely that the SM Higgs potential is metastable and unable to support Higgs Inflation.

\begin{figure}[h]
\begin{center}
\hspace*{-0.5cm}\includegraphics[trim = -3cm 0cm 0cm 0cm, clip = true, width=0.55\textwidth, angle = -90]{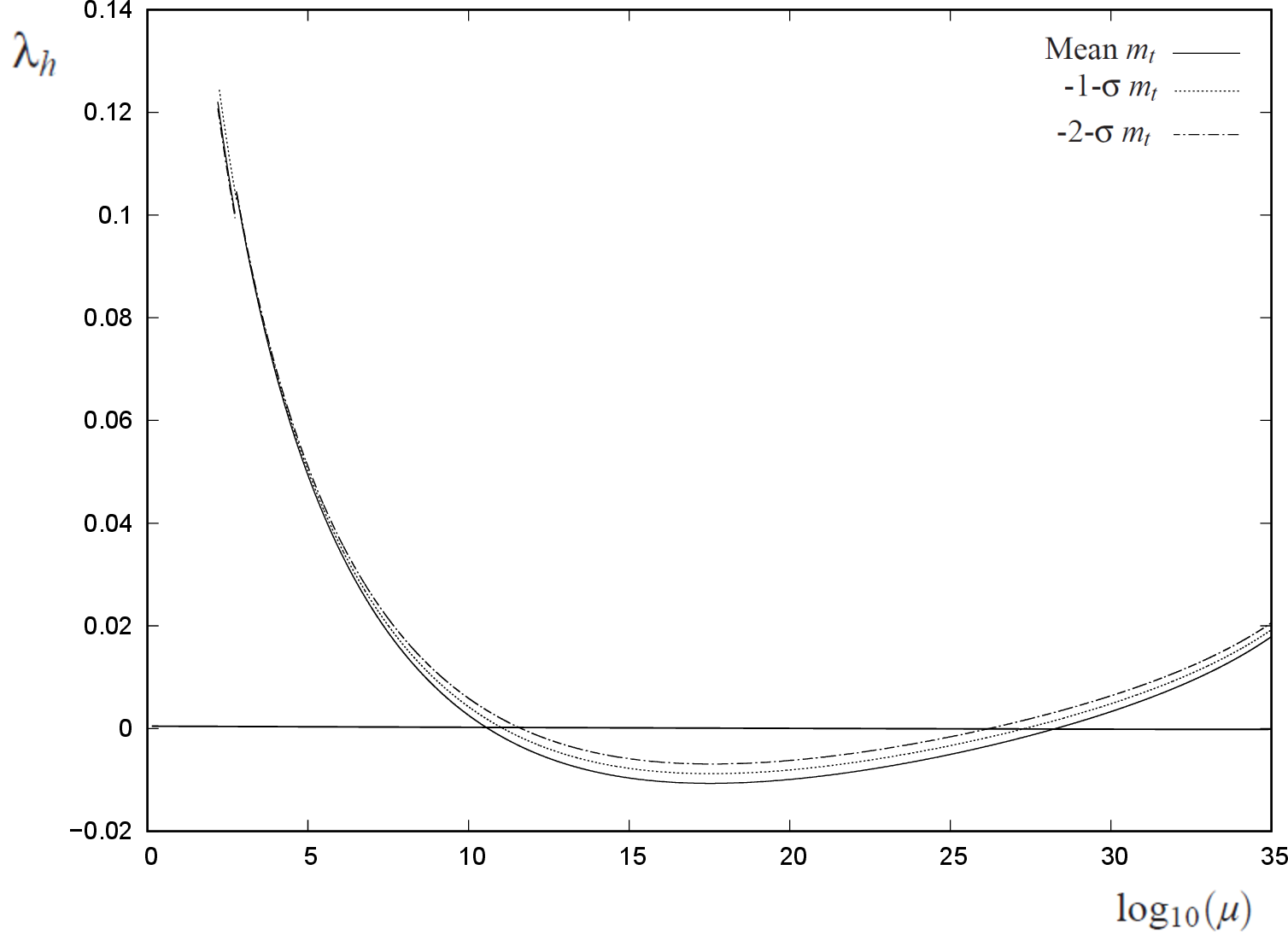}
\caption{SM Higgs coupling $\lambda_{h}(\mu)$ at $\mu = \phi$ for the mean, -1-$\sigma$ and -2-$\sigma$ values of $m_{t}$.} 
\label{fig1}
\end{center}
\end{figure}

\begin{figure}[h]
\begin{center}
\hspace*{-0.5cm}\includegraphics[trim = -3cm 0cm 0cm 0cm, clip = true, width=0.55\textwidth, angle = -90]{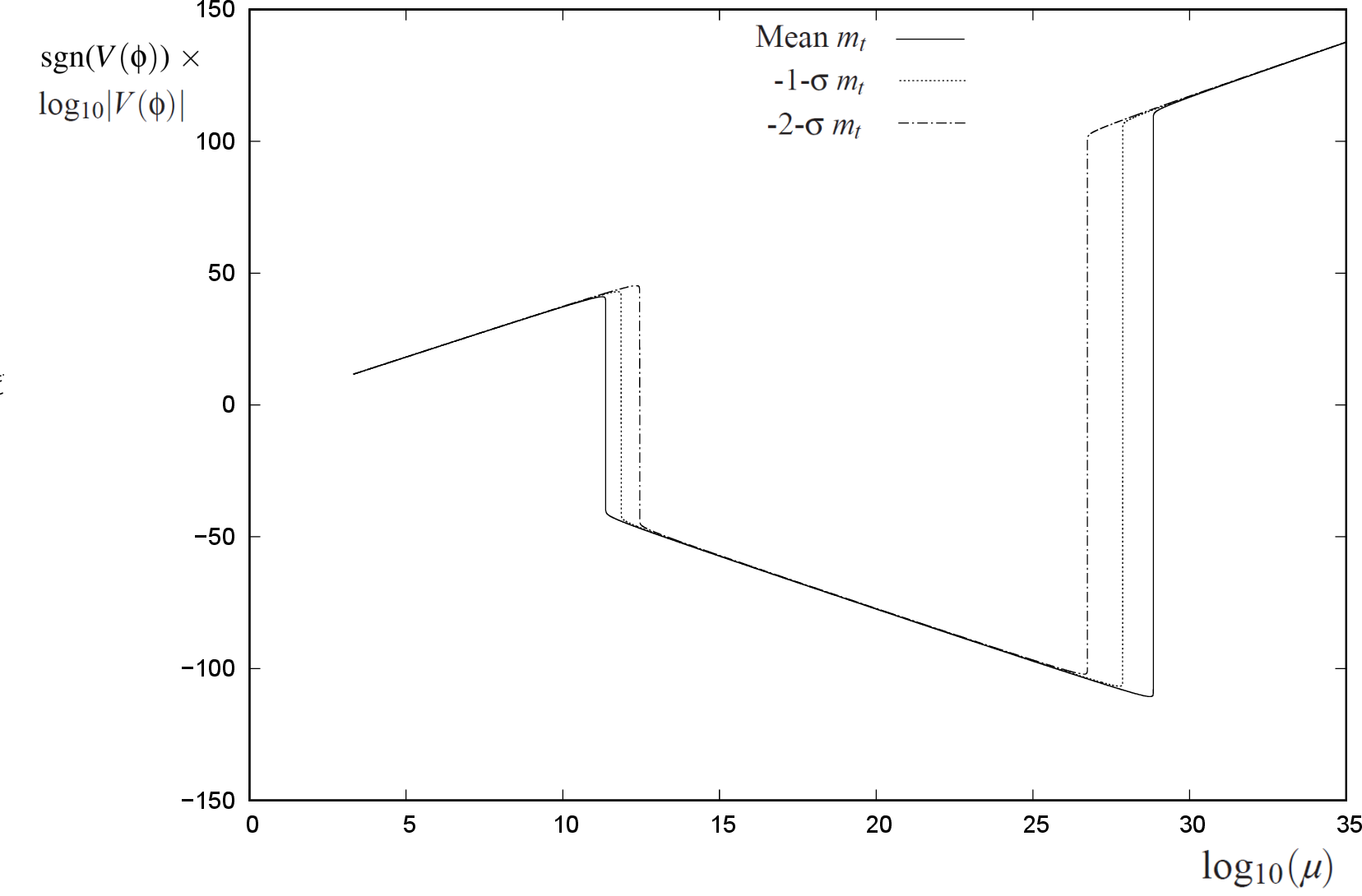}
\caption{SM Higgs potential calculated at $\mu = \phi$ for the mean, -1-$\sigma$ and -2-$\sigma$ values of $m_{t}$.} 
\label{fig2}
\end{center}
\end{figure}

\subsection{Quantum corrections in VLQ-stabilised Higgs Inflation} 

We again use the 3-loop ${\rm \overline{MS}}$ RG equations for the SM given in \cite{butt}, now modified to take into account the Higgs propagator suppression due to the kinetic term mixing of the physical Higgs boson with the graviton in the presence of a background $\phi$, which suppresses the physical Higgs boson propagator by a factor $s(\phi)$ \cite{wilczek, star}, 
\be{e5} s(\phi) = \frac{1 + \frac{\xi \phi^{2}}{M_{P}^{2}} }{1 + \left(6 \xi + 1 \right)\frac{\xi \phi^{2}}{M_{P}^{2}} } ~.\ee 
This results in a suppression of the contribution of the physical Higgs to the RG equations, but does not affect the Goldstone contribution. We have included the $s(t)$ factors at 1-loop given in \cite{star,corr1} and at 2-loop given in \cite{wilczek}. In practice, the propagator suppression has a small effect on the RG evolution of the couplings and Higgs potential.  Finally, we supplement the 3-loop SM RG equations with the 1-loop and leading 2-loop VLQ corrections given in \cite{vlqs1}, which we summarise in Appendix A. The $T$ and $B$ vector quarks will also have unknown Yukawa couplings to the SM quarks and Higgs boson. We will assume that these couplings are small enough to not significantly modify the Higgs potential.

\begin{figure}[h]
\begin{center}
\hspace*{-0.5cm}\includegraphics[trim = -3cm 0cm 0cm 0cm, clip = true, width=0.55\textwidth, angle = -90]{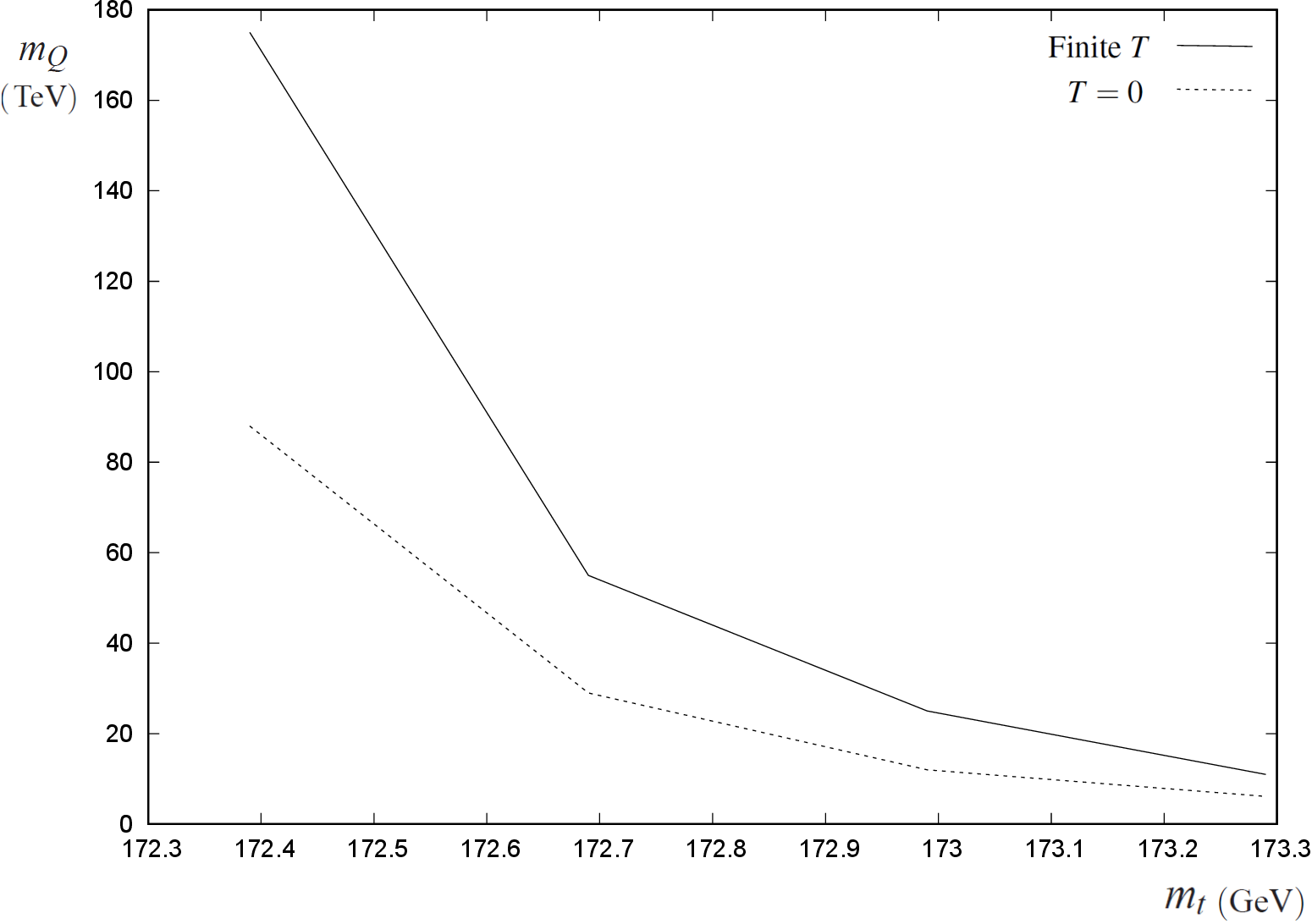}
\caption{Upper bound on $m_{Q}$ as a function of $m_{t}$ for $T$ vector quarks with $n_{Q} = 3$, from absolute stability and from stability of the finite temperature effective potential.} 
\label{fig1}
\end{center}
\end{figure} 

\begin{figure}[h]
\begin{center}
\hspace*{-0.5cm}\includegraphics[trim = -3cm 0cm 0cm 0cm, clip = true, width=0.55\textwidth, angle = -90]{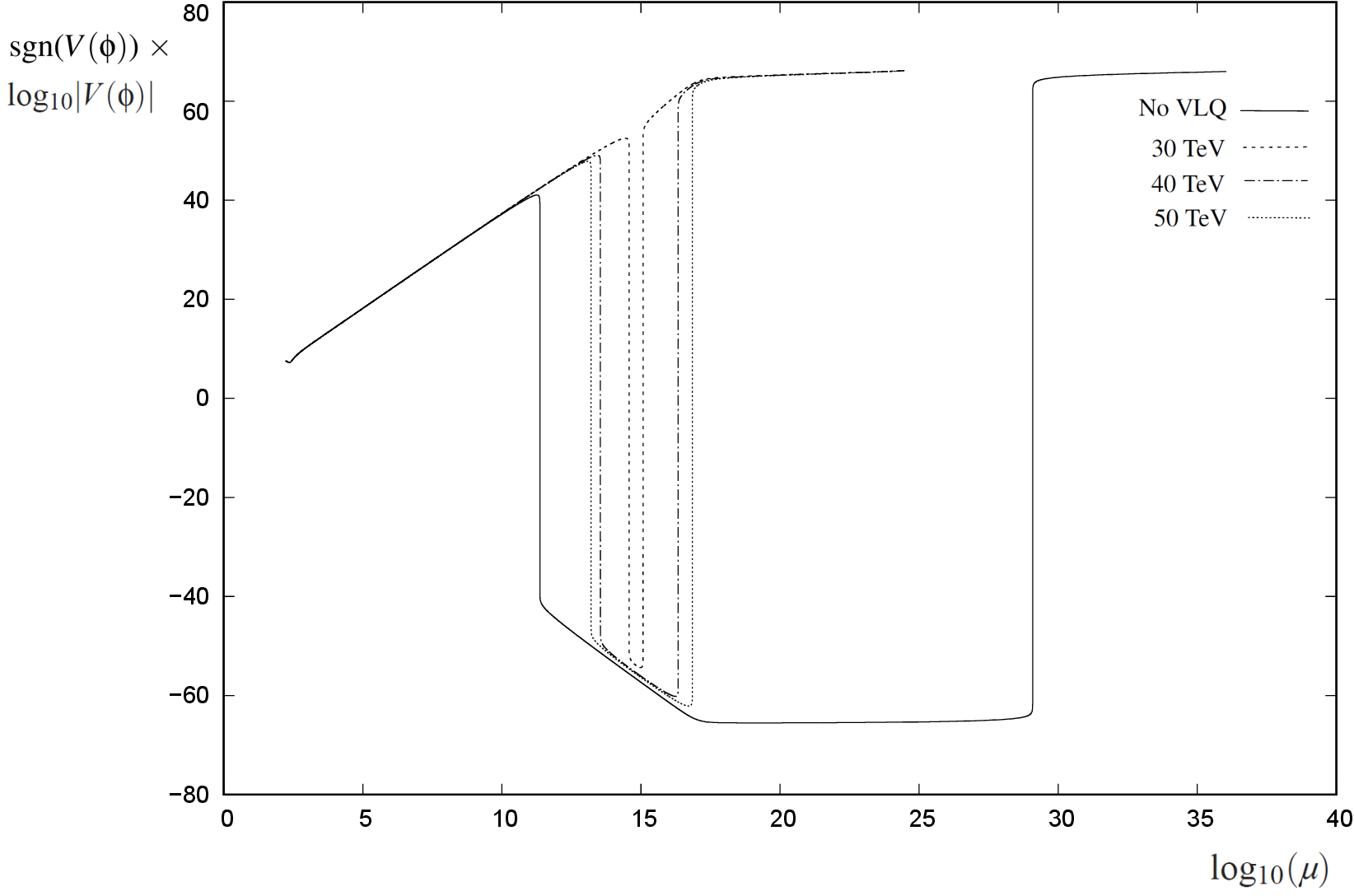}
\caption{The $T = 0$ Prescription II Higgs potential as a function of $m_{Q}$,  for the case of $T$ vector quarks with $n_{Q} = 3$ and mean $m_{t}$. The negative potential gap progressively closes as $m_{Q}$ is reduced.} 
\label{fig1}
\end{center}
\end{figure}

\begin{figure}[h]
\begin{center}
\hspace*{-0.5cm}\includegraphics[trim = -3cm 0cm 0cm 0cm, clip = true, width=0.55\textwidth, angle = -90]{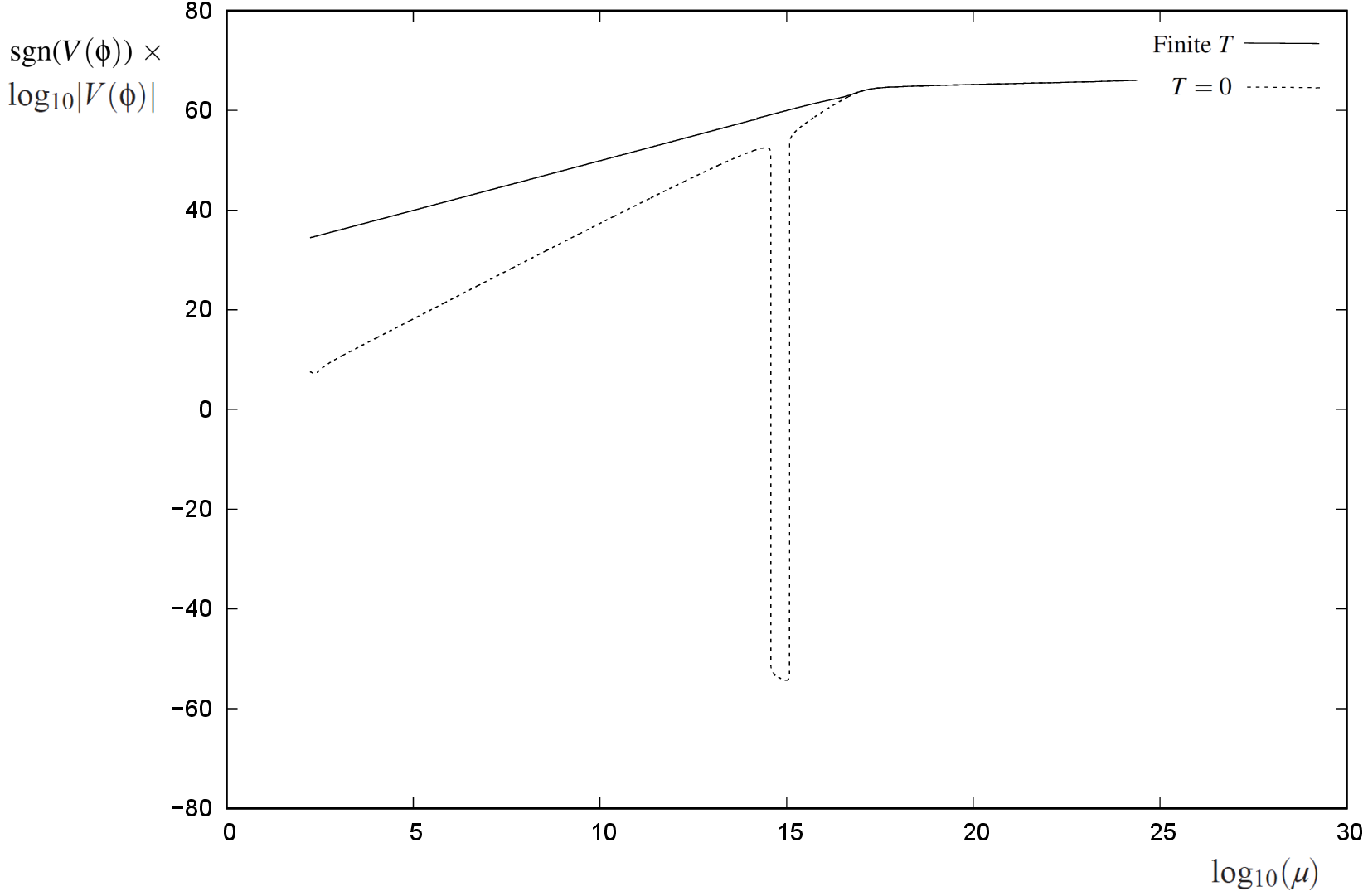}
\caption{Quantum corrected $T = 0$ potential and finite temperature effective potential, for $T$ vector quarks with $n_{Q} = 3$, $m_{Q} = 30 \GeV$ and mean $m_{t}$.} 
\label{fig1}
\end{center}
\end{figure}

\begin{figure}[h]
\begin{center}
\hspace*{-0.5cm}\includegraphics[trim = -3cm 0cm 0cm 0cm, clip = true, width=0.55\textwidth, angle = -90]{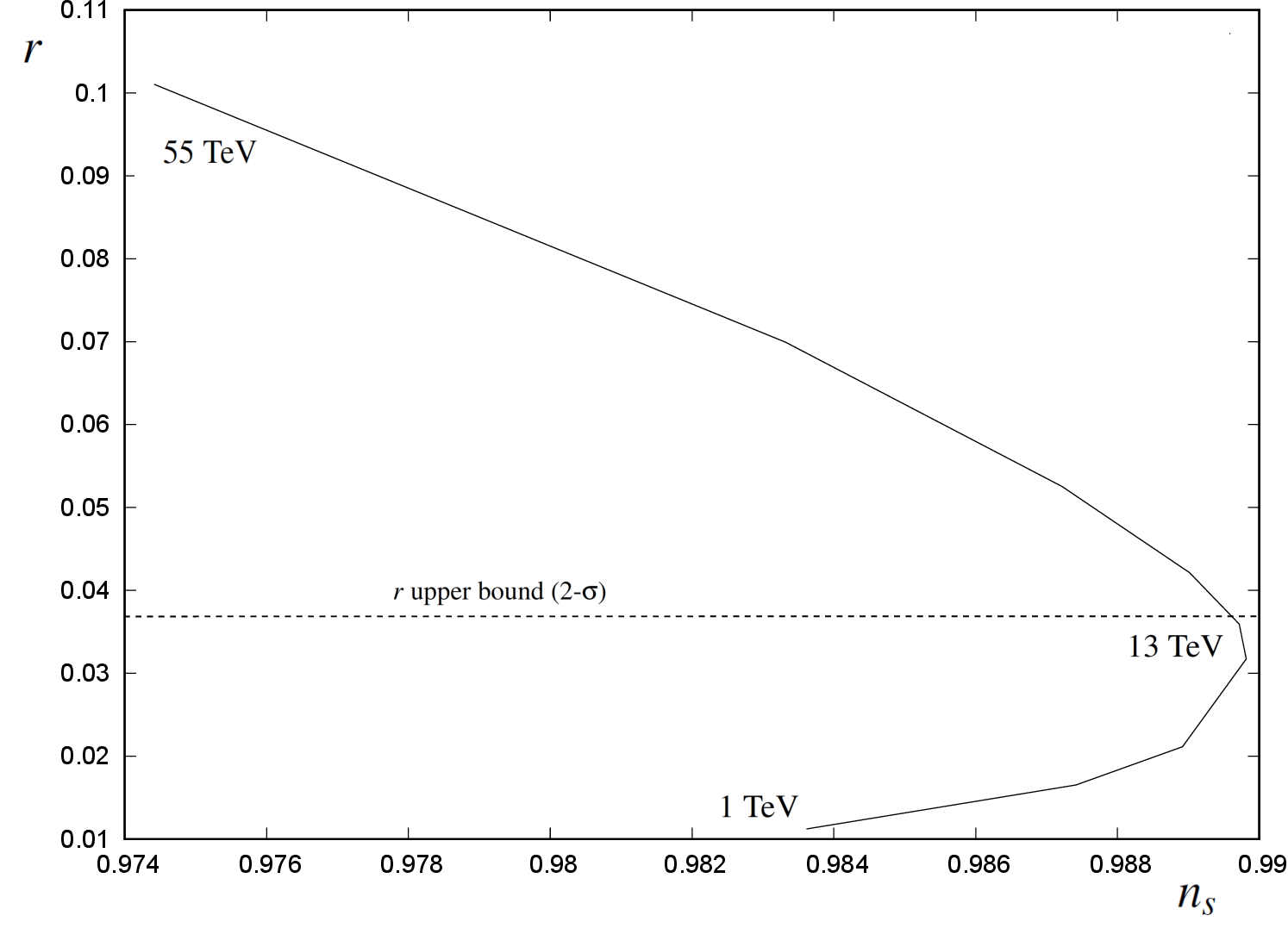}
\caption{Prescription II values of $r$ and $n_{s}$ as $m_{Q}$ varies from 1 TeV to 55 TeV, for the case of $T$ quarks witn $n_{Q} = 3$ and mean $m_{t}$. The observational upper bound on $r$ imposes an upper bound on $m_{Q}$ of around 13 TeV. The range of values of $n_{s}$ is 0.984 to 0.990 as $m_{Q}$ increases from 1 TeV to  13 TeV, with the smallest value of $r$ being 0.011.} 
\label{fig1}
\end{center}
\end{figure}

\begin{figure}[h]
\begin{center}
\hspace*{-0.5cm}\includegraphics[trim = -3cm 0cm 0cm 0cm, clip = true, width=0.55\textwidth, angle = -90]{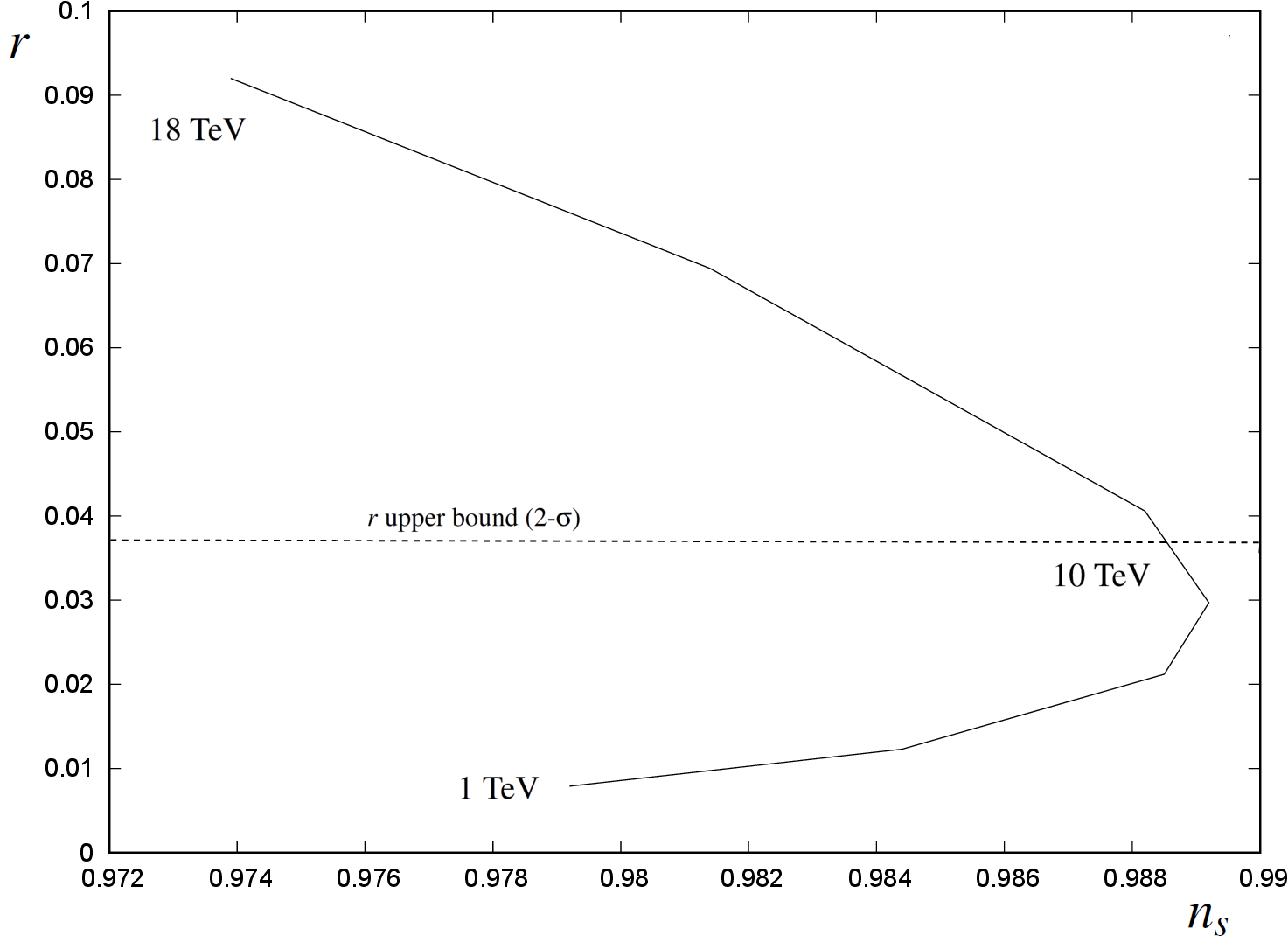}
\caption{Prescription II values of $r$ and $n_{s}$ as $m_{Q}$ varies from 1 TeV to 18 TeV, for the case of $B$ quarks with $n_{Q} = 3$ and mean $m_{t}$. The observational upper bound on $r$ imposes an upper bound on $m_{Q}$ of around 10 TeV. The range of values of $n_{s}$ is 0.979 to 0.989 as $m_{Q}$ increases from 1 TeV to 10 TeV. with the smallest value of $r$ being 0.0079.} 
\label{fig1}
\end{center}
\end{figure}

\begin{figure}[h]
\begin{center}
\hspace*{-0.5cm}\includegraphics[trim = -3cm 0cm 0cm 0cm, clip = true, width=0.55\textwidth, angle = -90]{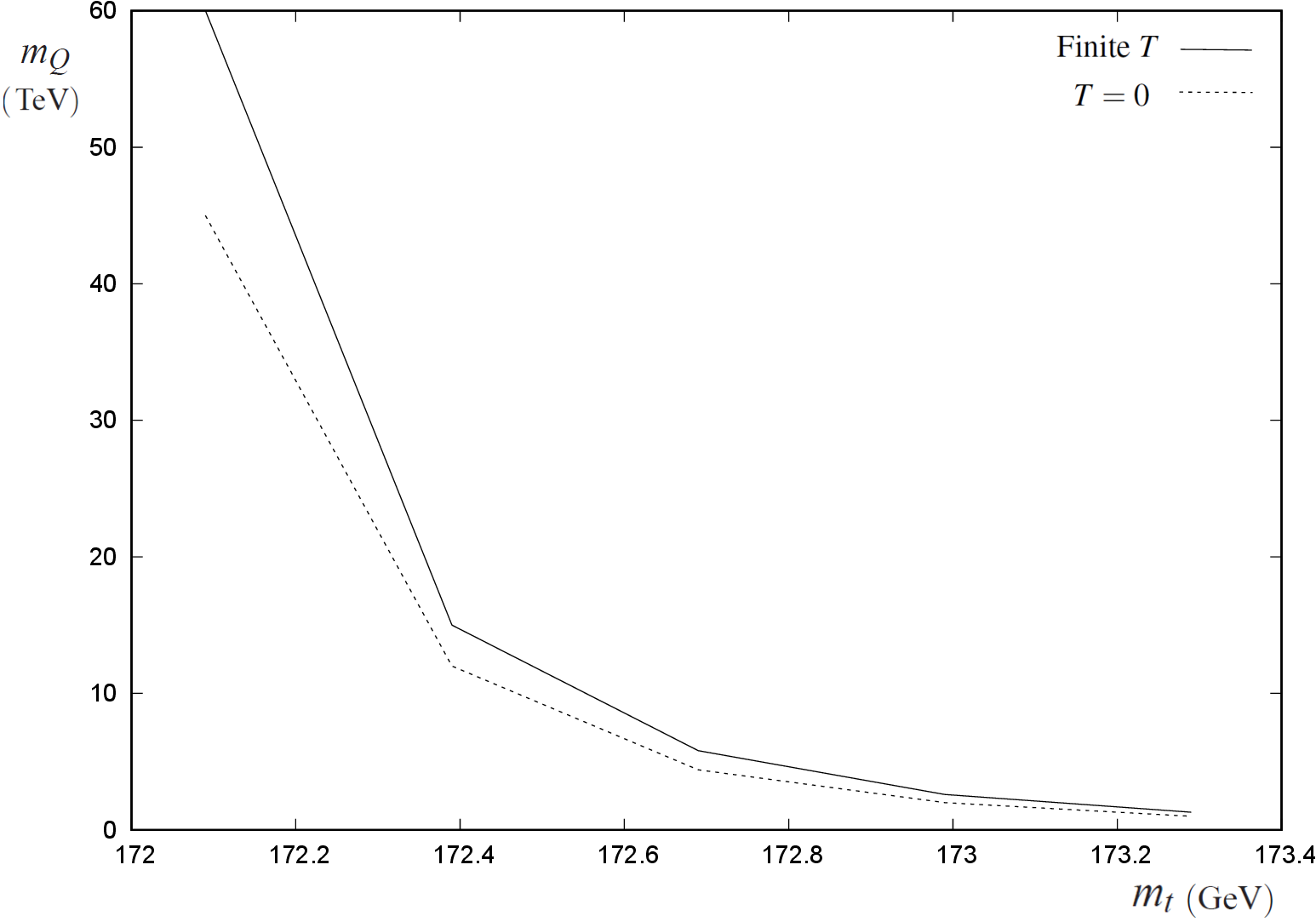}
\caption{Upper bound on $m_{Q}$ as a function of $m_{t}$ for $T$ vector quarks with $n_{Q} = 2$, from absolute stability and from stability of the finite temperature effective potential.} 
\label{fig1}
\end{center}
\end{figure}

\begin{figure}[h]
\begin{center}
\hspace*{-0.5cm}\includegraphics[trim = -3cm 0cm 0cm 0cm, clip = true, width=0.55\textwidth, angle = -90]{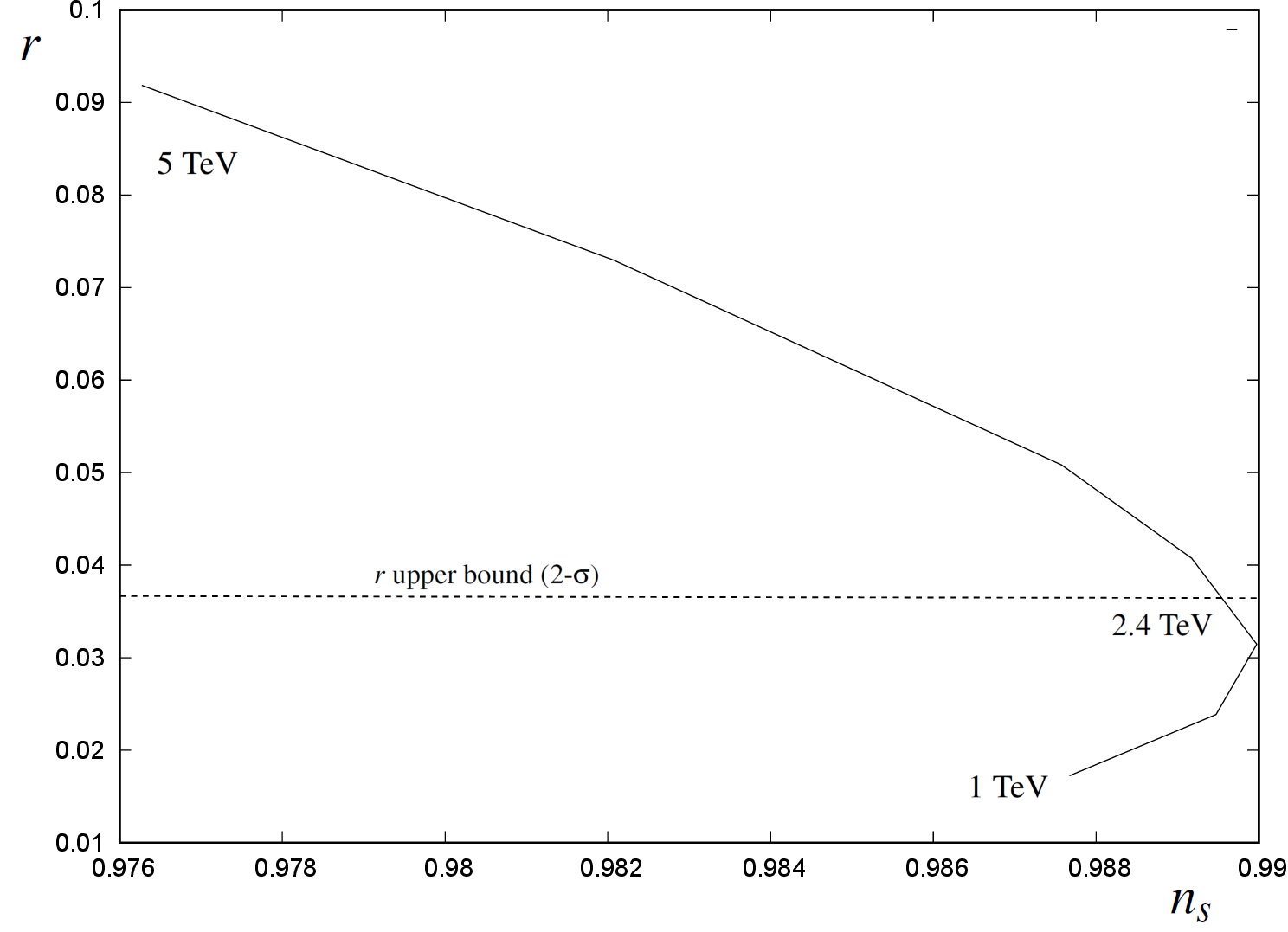}
\caption{Prescription II values of $r$ and $n_{s}$ as $m_{Q}$ varies from 1 TeV to 5 TeV, for the case of $T$ quarks with $n_{Q} = 2$ and mean $m_{t}$. The observational upper bound on $r$ imposes an upper bound on $m_{Q}$ of around 2.4 TeV. The range of values of $n_{s}$ is 0.988 to 0.990 as $m_{Q}$ increases from 1 TeV to 2.4 TeV, with the smallest value of $r$ being 0.017.} 
\label{fig1}
\end{center}
\end{figure}

\begin{figure}[h]
\begin{center}
\hspace*{-0.5cm}\includegraphics[trim = -3cm 0cm 0cm 0cm, clip = true, width=0.55\textwidth, angle = -90]{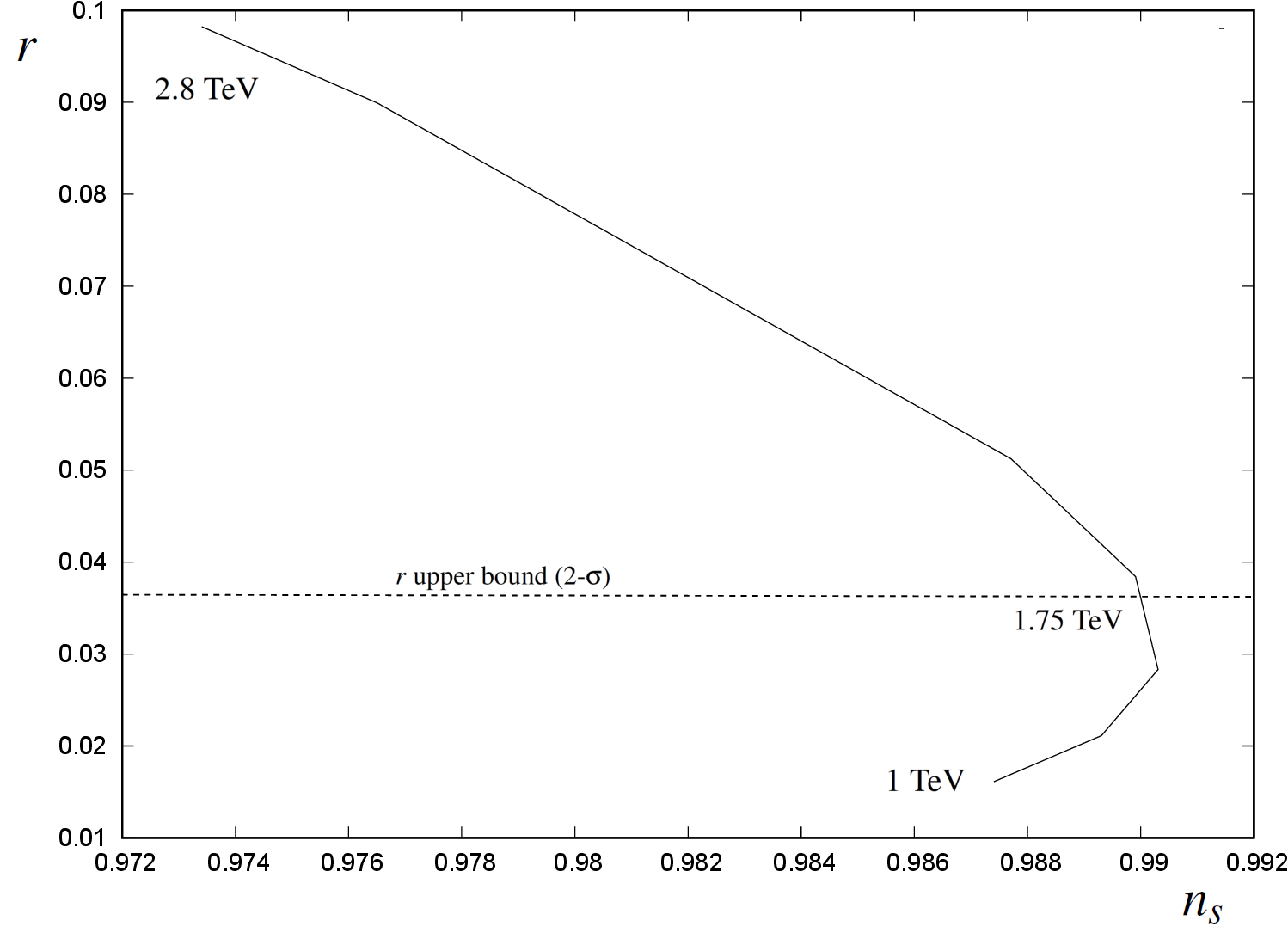}
\caption{Prescription II values of $r$ and $n_{s}$ as $m_{Q}$ varies from 1 TeV to 2.8 TeV, for the case of $B$ quarks with $n_{Q} = 2$ and mean $m_{t}$. The observational upper bound on $r$ imposes an upper bound on $m_{Q}$ of around 1.75 TeV. The range of values of $n_{s}$ is 0.987 to 0.990 as $m_{Q}$ increases from 1 TeV to 1.75 TeV. with the smallest value of $r$ being 0.015.} 
\label{fig1}
\end{center}
\end{figure}

\subsection{Renormalisation prescription} 

The form of the quantum correction to the potential during inflation depends upon the conformal frame in which the model is renormalised. Two frames are commonly considered, known as Prescription I, where the model is renormalised in the Einstein frame, and Prescription II,  where the model is renormalised in the Jordan frame and the complete quantum corrected Higgs potential is transformed to the Einstein frame \cite{presc1,presc2}. These are in effect two completely different versions of Higgs Inflation, which correspond to different UV completions of the model \cite{uvframe}. In the case of Prescription II, we will run the SM RG equations (including the Higgs propagator suppression) up to $\mu = \phi$. In the case of Prescription I, the SM RG equations become invalid once the conformal factor significantly deviates from 1 and $\mu = \phi \gae \phi_{c}$. We will therefore run to RG equations up to $\mu_{c} = 0.1 \phi_{c}$ and then use the 1-loop CW correction to compute the potential at $\phi > \phi_{c}$. Since the quantum corrections become rapidly independent of $\phi$ at $\phi > \phi_{c}$ in Prescription I, this choice of renormalisation scale is large enough to prevent large logarithms in the 1-loop CW potential.     

\subsection{Perturbative unitarity violation}

Higgs scalar scattering in the cosmological background during inflation is expected to violate unitarity at energies $E \approx \phi$, corresponding to the mass of the $W$ and $Z$ bosons during inflation, where $\phi$ is the background Higgs field during inflation \cite{hiuv,hiuv2}. If perturbative unitarity indicates true unitarity violation, then a change in the theory is necessary, characterised by the energy scale $\phi$. In this case it is possible that new physics could modify the effective potential at the scale $\phi$ \footnote{In \cite{rr1, rr2} it is proposed that the new physics associated with the UV completion could introduce threshold effects that modify the SM couplings at the scale $\mu \sim M_{Pl}/\xi$, This could alter the Higgs potential and allow for inflation even if the unmodified SM Higgs potential is unstable.}.  Alternatively, perturbative unitarity violation in Higgs scattering may instead indicate a breakdown of perturbation theory, with unitarity conserved non-perturbatively \cite{sh1,sh2,sh3,sh4}. In this case no new physics is necessary to conserve unitarity, and since the effect of the non-minimal coupling on the RG equations and the effective potential is taken into account by the Higgs propagator suppression, there is no reason for the effective potential to be modified. In the following we assume that any corrections associated with unitarity conservation are small enough to be neglected.

\subsection{The 1-loop Coleman-Weinberg correction} 

To calculate the quantum corrected Higgs potential, we run the couplings up to the renormalisation scale $\mu$ and then add the 1-loop CW correction. In general, the
${\rm \overline{MS}}$ scheme SM 1-loop CW potential is given by \cite{sher}
\be{e6} \Delta V_{1-loop} = \sum_{i} \;\frac{C_{i} M_{i}^{4}}{64 \pi^{2}} \left[\ln\left(\frac{M_{i}^{2}}{\mu^{2}}\right) - K_{i}\right]   ~,\ee
where $M_{i}$ the mass in the renormalisation frame and $(C_{i}, K_{i})  = (3, 3/2)$ for the Goldstone bosons, (6, 5/6) for the $W$ bosons, (3, 5/6) for the $Z$ boson, and (-12, 3/2) for the t-quark. In these we have summed over the 3 Goldstone bosons, 2 $W$ bosons and all $t$-quark colours. 
We do not include the physical SM Higgs boson as its contribution is suppressed by the non-minimal coupling propagator suppression. The particle masses are $M_{i} = M_{J,\,i}$ in the Jordan frame and $M_{J,\,i}/\Omega$ in the Einstein frame. The terms in the 1-loop CW potential in the chosen renormalisation frame therefore have the form   
\be{e7} \Delta V_{1-loop} \sim \frac{C \,M_{J}^{4}}{16 \pi^{2} \Omega^{4 \alpha}} \left( \ln \left(\frac{M_{J}^{2}(\phi)}{\mu^{2} \Omega^{2 \alpha} }\right) - K \right)  ~, \ee 
where $\alpha = 1$ for Prescription I and $\alpha = 0$ for Prescription II. Once transformed the Einstein frame, the final 1-loop potential is of the form
\be{e8} V_{E}(\phi) = \frac{V(\phi)}{\Omega^{4}}  + \sum_{i} \frac{C_{i}\; M_{J,\,i}^{4}}{16 \pi^{2} \Omega^{4}} \left[ \ln \left(\frac{M_{J,\,i}^{2}(\phi)}{\mu_{c}^{2} \Omega^{2 \alpha} }\right) - K_{i} \right]  ~.\ee

\section{The finite temperature effective potential} 

The condition for successful Higgs Inflation is that the Higgs expectation value can evolve into the electroweak vacuum. This does not require absolute stability but only that the minimum of the finite temperature effective potential (FTEP) after inflation, which gives the thermal equilibrium expectation value of the Higgs field, is at $\phi = 0$. We refer to this as
 finite temperature stability. Due to the large couplings of the Higgs boson to the SM fields, after inflation the Higgs field oscillations rapidly preheat and decay to SM fields \cite{hbb,rubio1,rubio2}. We will therefore assume that reheating is instantaneous and compute the FTEP after inflation to see if it is stable. We calculate the FTEP in the Einstein frame,  which is the frame in which inflation is analysed, so the mass terms entering the FTEP are calculated in the Einstein frame. These are generally related to the conventional (i.e. Jordan frame) SM masses $M_{J,\,i}$ by $M_{i} = M_{J,\,i}/\Omega$. In practice, the Einstein frame is essentially the same as the Jordan frame when calculating the FTEP, since $\Omega \approx 1$ after inflation and reheating.

The FTEP is given by $V_{E}(\phi, T) = V_{E}(\phi) + \Delta V(\phi, T)$, where 
\be{e9} \Delta V(\phi, T) = \frac{T^{4}}{2 \pi^{2}} \sum_{i} n_{B,\, i} \, I_{B}\left[ \frac{M_{i}^{2}(\phi)}{T^{2}} \right]  -  \frac{T^{4}}{2 \pi^{2}} \sum_{i} n_{F,\,i} \, I_{F}\left[ \frac{M_{i}^{2}(\phi)}{T^{2}} \right]   ~\ee 
where the integrals $I_{B}$ and $I_{F}$ are given by
\be{e10} I_{B}\left[\frac{M_{i}^{2}(\phi)}{T^{2}} \right] = 
\int_{0}^{\infty} dx\, x^{2} \ln\left[1 - \exp\left( - \sqrt{x^2 + \frac{M_{i}^{2}(\phi)}{T^{2}}} \right)  \right]  ~\ee 
and
\be{e11} I_{F}\left[\frac{M_{i}^{2}(\phi)}{T^{2}} \right] = 
\int_{0}^{\infty} dx\, x^{2} \ln\left[1 + \exp\left( - \sqrt{x^2 + \frac{M_{i}^{2}(\phi)}{T^{2}}} \right)  \right]  ~,\ee
where $n_{B,\,i}$ and $n_{F,\,i}$ are the number of degrees of freedom of the bosons and fermions. 
In the figures we show $V_{E}(\phi, T) - \Delta V(0, T)$ so that the finite temperature contribution equals zero at $\phi = 0$.

\section{Results} 

The potential during inflation depends upon the value of the non-minimal coupling. This is determined by requiring that the curvature perturbation power spectrum at the pivot scale is equal to its observed value. The number of e-foldings at the pivot scale is discussed in Appendix B. We find that in general the potential at the end of inflation is $V_{end} \approx 10^{64} \GeV^{4}$, where $V_{end} = V_{E}(\sigma_{end})$ is the Einstein frame potential at the end of inflation, and so $N_{*} \approx 57$ for the case of instant reheating. 

The present experimental lower bound on the mass of vector quarks is around 1 TeV, with the exact lower bound depending on the assumed decay mode of the vector quark \cite{atlas,cms,bel,fc}. We will therefore impose a lower bound of 1 TeV on $m_{Q}$ in our results.

\begin{table}[htbp]
\begin{center}
\begin{tabular}{ |c|c|c| }
\hline
$m_{t} ({\rm GeV})$ & $m_{Q}$ (${\rm \, Abs}$) 
& $m_{Q}$ (${\rm \, Finite\;T}$)   
\\
\hline
173.29 ($+$2-$\sigma$) 
& $6.1 \TeV$ 
& $11 \TeV$ 
\\
172.99 ($+$1-$\sigma$) 
& $12 \TeV$ 
&  $25 \TeV$ 
\\
172.69 (mean) 
& $29 \TeV$ 
& $55 \TeV$ 
\\
172.39 ($-$1-$\sigma$) 
& $88 \TeV$ 
& $175 \TeV$ 
\\
172.09 ($-$2-$\sigma$) 
& $380 \TeV$ 
& $680 \TeV$ 
\\
\hline
\end{tabular}
\caption{$m_{Q}$ upper bound as a function of $m_{t}$ for $T$ vector quarks with $n_{Q}$ = 3, from absolute and 
finite temperature stability.  }
\end{center}
\end{table}

\begin{table}[htbp]
\begin{center}
\begin{tabular}{ |c|c|c|c|c| }
\hline
$m_{Q}$ 
& $n_{s}$ 
& $r$ & $\xi(m_{t})/10^{3}$ & $V_{end} \,(\GeV^4)$ 
\\
\hline
$55 {\rm \, TeV} $ 
& $0.9744$ 
& $0.101$ 
& $0.45$
& $1.31  \times 10^{64}$
\\
$29 {\rm \, TeV}$ 
& $0.9833$ 
& $6.99 \times 10^{-2}$ 
& $0.58$
& $1.89 \times 10^{64}$
\\
$20 {\rm \, TeV} $ 
& $0.9872$ 
& $5.25 \times 10^{-2}$ 
& $0.70$
& $1.86 \times 10^{64}$
\\
$15 {\rm \, TeV}$ 
& $0.9890$ 
& $4.21 \times 10^{-2}$ 
& $0.84$
& $1.63 \times 10^{64}$
\\
$13 {\rm \, TeV}$ 
& $0.9895$ 
& $3.80 \times 10^{-2}$ 
& $0.91$
& $1.53 \times 10^{64}$
\\
$12 {\rm \, TeV}$ 
& $0.9897$ 
& $3.59 \times 10^{-2}$ 
& $0.93$
& $1.55 \times 10^{64}$
\\
$10 {\rm \, TeV}$ 
& $0.9898$ 
& $3.17 \times 10^{-2}$ 
& $1.05$
& $1.37 \times 10^{63}$
\\
$5 {\rm \, TeV}$ 
& $0.9889$ 
& $2.11 \times 10^{-2}$ 
& $1.45$
& $1.05 \times 10^{63}$
\\
$3 {\rm \, TeV}$ 
& $0.9874$ 
& $1.65 \times 10^{-2}$ 
& $1.84$
& $8.31 \times 10^{63}$
\\

$1 {\rm \, TeV}$ 
& $0.9836$ 
& $1.12 \times 10^{-2}$ 
& $2.70$
& $6.07 \times 10^{63}$
\\
\hline
\end{tabular}
\caption{Prescription II inflation observables and parameters as a function of $m_{Q}$, for $T$ vector quarks with $n_{Q} = 3$ and mean $m_{t}$.}
\end{center}
\end{table}

\begin{table}[htbp]
\begin{center}
\begin{tabular}{ |c|c|c|c|c| }
\hline
$m_{Q}$ 
& $n_{s}$ 
& $r$ & $\xi(m_{t})/10^{3}$ & $V_{end}\,(\GeV^4)$ 
\\
\hline
$18 {\rm \, TeV} $ 
& $0.9739$ 
& $9.20 \times 10^{-2}$ 
& $0.35$
& $1.15  \times 10^{64}$
\\
$14 {\rm \, TeV}$ 
& $0.9814$ 
& $6.94 \times 10^{-2}$ 
& $0.42$
& $1.70 \times 10^{64}$
\\
$10 {\rm \, TeV} $ 
& $0.9882$ 
& $4.06 \times 10^{-2}$ 
& $0.60$
& $1.60 \times 10^{64}$
\\
$9 {\rm \, TeV} $ 
& $0.9890$ 
& $3.48 \times 10^{-2}$ 
& $0.68$
& $1.44 \times 10^{64}$
\\
$8 {\rm \, TeV}$ 
& $0.9892$ 
& $2.97 \times 10^{-2}$ 
& $0.75$
& $1.36 \times 10^{64}$
\\
$6 {\rm \, TeV}$ 
& $0.9885$ 
& $2.12 \times 10^{-2}$ 
& $1.02$
& $1.00 \times 10^{64}$
\\
$3 {\rm \, TeV}$ 
& $0.9844$ 
& $1.23 \times 10^{-2}$ 
& $1.65$
& $6.61 \times 10^{63}$
\\
$1 {\rm \, TeV}$ 
& $0.9792$ 
& $7.87 \times 10^{-3}$ 
& $2.70$
& $4.56 \times 10^{63}$
\\
\hline
\end{tabular}
\caption{Prescription II inflation observables and parameters as a function of $m_{Q}$, for $B$ vector quarks with $n_{Q} = 3$  and mean $m_{t}$.}
\end{center}
\end{table}

\begin{table}[htbp]
\begin{center}
\begin{tabular}{ |c|c|c|c|c| }
\hline
$m_{t} ({\rm GeV})$ & $n_{s}$
& $r$ & $\xi(m_{t})/10^{3}$ & $V_{end}\,(\GeV^4)$   
\\
\hline
173.29 ($+$2-$\sigma$) 
& 0.9896
& $3.24 \times 10^{-2}$
& 1.07
& $1.43 \times 10^{64}$ GeV
\\
172.99 ($+$1-$\sigma$) 
& 0.9890
& $2.21 \times 10^{-2}$
& 1.46
& $1.04 \times 10^{64}$ GeV
\\
172.69 (Mean) 
& 0.9873
& $1.65 \times 10^{-2}$
& 1.84
& $8.31 \times 10^{63}$ GeV
\\
172.39 ($-$1-$\sigma$) 
& 0.9854
& $1.34 \times 10^{-2}$
& 2.21
& $6.98 \times 10^{63}$ GeV
\\
172.09 ($-$2-$\sigma$) 
& 0.9839
& $1.12 \times 10^{-2}$
& 2.57
& $6.08 \times 10^{63}$ GeV
\\
\hline
\end{tabular}
\caption{Prescription II inflation observables and parameters as a function of $m_{t}$, for $T$ vector quarks with $n_{Q} = 3$ and $m_{Q} =$ 3 TeV.  }
\end{center}
\end{table}

\begin{table}[htbp]
\begin{center}
\begin{tabular}{ |c|c|c| }
\hline
$m_{t} ({\rm GeV})$ & $m_{Q}$ (${\rm \, Abs}$) 
& $m_{Q}$ (${\rm \, Finite\;T}$)   
\\
\hline
173.29 ($+$2-$\sigma$) 
& $1.0 \TeV$ 
& $1.3 \TeV$ 
\\
172.99 ($+$1-$\sigma$) 
& $2.0 \TeV$ 
&  $2.3 \TeV$ 
\\
172.69 (mean) 
& $4.4 \TeV$ 
& $5.8 \TeV$ 
\\
172.39 ($-$1-$\sigma$) 
& $12 \TeV$ 
& $15 \TeV$ 
\\
172.09 ($-$2-$\sigma$) 
& $45 \TeV$ 
& $60 \TeV$ 
\\
\hline
\end{tabular}
\caption{$m_{Q}$ upper bound as a function of $m_{t}$, for $T$ vector quarks with $n_{Q} = 2$, from absolute and finite temperature stability.  }
\end{center}
\end{table}

\begin{table}[htbp]
\begin{center}
\begin{tabular}{ |c|c|c|c|c| }
\hline
$m_{Q}$ 
& $n_{s}$ 
& $r$ & $\xi(m_{t})/10^{3}$ & $V_{end}\,(\GeV^4)$ 
\\
\hline
$5.0 {\rm \, TeV} $ 
& $0.9763$ 
& $9.14 \times 10^{-2}$ 
& $0.40$
& $1.46  \times 10^{64}$
\\
$4.0 {\rm \, TeV}$ 
& $0.9821$ 
& $7.25 \times 10^{-2}$ 
& $0.46$
& $1.84 \times 10^{64}$
\\
$3.0 {\rm \, TeV} $ 
& $0.9876$ 
& $5.04 \times 10^{-2}$ 
& $0.59$
& $1.76 \times 10^{64}$
\\
$2.4 {\rm \, TeV}$ 
& $0.9895$ 
& $3.84 \times 10^{-2}$ 
& $0.72$
& $1.55 \times 10^{64}$
\\
$2.0 {\rm \, TeV}$ 
& $0.9900$ 
& $3.10 \times 10^{-2}$ 
& $0.85$
& $1.33 \times 10^{64}$
\\
$1.5 {\rm \, TeV}$ 
& $0.9895$ 
& $2.34 \times 10^{-2}$ 
& $1.06$
& $1.10 \times 10^{64}$
\\
$1.0 {\rm \, TeV}$ 
& $0.9877$ 
& $1.68 \times 10^{-2}$ 
& $1.40$
& $8.54 \times 10^{63}$
\\
\hline
\end{tabular}
\caption{Prescription II inflation observables and parameters as a function of $m_{Q}$, for $T$ vector quarks with $n_{Q} = 2$ and mean $m_{t}$. }
\end{center}
\end{table}

\begin{table}[htbp]
\begin{center}
\begin{tabular}{ |c|c|c|c|c| }
\hline
$m_{Q}$ 
& $n_{s}$ 
& $r$ & $\xi(m_{t})/10^{3}$ & $V_{end}\,(\GeV^4)$ 
\\
\hline
$2.8 {\rm \, TeV} $ 
& $0.9734$ 
& $9.76 \times 10^{-2}$ 
& $0.29$
& $1.28 \times 10^{64}$
\\
$2.6 {\rm \, TeV}$ 
& $0.9765$ 
& $8.93 \times 10^{-2}$ 
& $0.31$
& $1.61 \times 10^{64}$
\\
$2.0 {\rm \, TeV} $ 
& $0.9877$ 
& $5.06 \times 10^{-2}$ 
& $0.45$
& $1.84 \times 10^{64}$
\\
$1.75 {\rm \, TeV}$ 
& $0.9899$ 
& $3.78 \times 10^{-2}$ 
& $0.56$
& $1.55 \times 10^{64}$
\\
$1.5 {\rm \, TeV}$ 
& $0.9903$ 
& $2.77 \times 10^{-2}$ 
& $0.85$
& $1.29 \times 10^{64}$
\\
$1.25 {\rm \, TeV}$ 
& $0.9893$ 
& $2.05 \times 10^{-2}$ 
& $0.71$
& $1.26 \times 10^{64}$
\\
$1.0 {\rm \, TeV}$ 
& $0.9874$ 
& $1.55 \times 10^{-2}$ 
& $1.15$
& $7.91 \times 10^{63}$
\\
\hline
\end{tabular}
\caption{Prescription II inflation observables and parameters as a function of $m_{Q}$, for $B$ vector quarks with $n_{Q} = 2$ and mean $m_{t}$.}
\end{center}
\end{table}

\subsection{${\bf n_{Q} = 3}$}

For the mean t-quark mass and three $T$ vector quarks, we find that the absolute stability bound is $m_{Q} \leq 29 \TeV$, and the finite temperature stability bound is $m_{Q} \leq 55 \TeV $. For $B$ vector quarks, the corresponding bounds are $m_{Q} \leq 14 \TeV$ and $m_{Q} \leq 18 \TeV$ respectively. 

In Table 1 and Figure 3 we show how the stability bounds vary with $m_{t}$ over the 2-$\sigma$ observed range for three $T$ quarks. The upper bounds on $m_{Q}$ are sensitive to $m_{t}$, becoming smaller as $m_{t}$ increases and the SM instability strengthens.

In Figure 4 we show the effect of decreasing $m_{Q}$ on the $T = 0$ potential and its stability for Prescription II with three $T$ vector quarks. For the case without VLQs, the upper bound on the gap is around $10^{29}$ GeV and the lower bound is around $10^{11} \GeV$. Once the vector quarks are introduced, the lower bound of the gap increases and the upper bound (more rapidly) decreases as $m_{Q}$ decreases, until the gap disappears and the potential becomes absolutely stable at $m_{Q} = 29 \GeV$.

In Figure 5 we show the $T = 0$ and FTEP potential for $m_{Q} = 30 $ TeV and three $T$ vector quarks, showing that the $T = 0$ potential is metastable, with a deep minimum at $\phi \neq 0$, whereas the FTEP has a minimum only at $\phi = 0$. In this case $\phi$ will cool into the electroweak vacuum after inflation,  even though the SM potential is metastable.

The stability upper bounds apply to both Prescription I and II, as $\phi_{upper} < \phi_{c}$ when stability breaks down and so the Jordan and Einstein frames are equivalent as far as stability of the potential is concerned.

We next consider how the inflation predictions vary as a function of $m_{Q}$. For Prescription I, we find that the inflation predictions are almost exactly equal to the classical Higgs Inflation predictions, $n_{s} = 0.966$ and $r = 3.2 \times 10^{-3}$, for all $m_{Q}$. This is due to the suppression of the Einstein frame mass terms by $\Omega$, which means that at $\phi > \phi_{c}$ the 1-loop CW corrections quickly becomes independent of $\phi$. 

For Prescription II, since the quantum corrections are calculated in the Jordan frame, the quantum corrections to the potential are not cut off at $\phi > \phi_{c}$.  In Figure 6 we show $r$ versus $n_{s}$ for three $T$ vector quarks for $m_{Q}$ in the range 1 TeV to 55 TeV and in Table 2 we show the predictions of the model for $n_{s}$, $r$, $\xi(m_{t})$ and $V_{end}$, with $m_{t}$ equal to its mean value.

We find from Figure 6 that the upper bound on $m_{Q}$ from the observational 2-$\sigma$ upper bound on $r$ is stronger than the bound from stability of the potential. Imposing the 2-$\sigma$ upper bound $r <  0.037$ \cite{rbound} gives an upper bound on $m_{Q}$ of 13 TeV for three $T$ quarks, compared to 55 TeV from finite temperature stability.      

For $1 \TeV \leq m_{Q} \leq 13$ TeV the range of $r$ is about 0.01 to 0.04. Therefore the predicted range of values for $r$ are not far below the present observational limits and will be easily observable by the next generation of CMB polarisation experiments, which will be able to observe $r$ down to $O(10^{-3})$ \cite{lite}. 

The values of $n_{s}$ for $1 \TeV \leq m_{Q} \leq 13$ TeV are in the range 0.984 to 0.990. These values are large compared to $\Lambda$CDM CMB range from Planck, 
$n_{s} = 0.965 \pm 0.004$ \cite{planck}. However, $\Lambda$CDM is now challenged by the $H_{0}$ tension between the $\Lambda$CDM CMB value of $H_{0}$ and the late-time supernova distance measurement of $H_{0}$. In the case of early-time solutions of the Hubble tension, where the tension is resolved by modifying the sound horizon at decoupling via an additional energy density component, in particular Early Dark Energy (EDE),  the value of $n_{s}$ from CMB is larger than the $\Lambda$CDM value \cite{taka,giare}. This can be understood as due to the additional scale dependent suppression of temperature fluctuations at smaller length scales due to the additional energy density component, which requires a larger $n_{s}$ to compensate. For the case of a typical EDE model based on an axion-like potential, the best-fit values are in the range 0.981-0.996, depending on the data set \cite{ede}. Therefore Prescription II Higgs Inflation with VLQ stabilisation can provide a minimal model for inflation based on TeV scale physics that is also naturally compatible with early-time solutions to the $H_{0}$ tension.

In Figure 7 and Table 3 we show the corresponding predictions for three $B$ quarks. In this case the observational bound on $r$ imposes an upper bound on $m_{Q}$ of 10 TeV, as compared to 18 TeV for finite temperature stability.  As $m_{Q}$ varies from 1 TeV to 10 TeV the value of $r$ varies from about 0.008 to 0.04, whilst $n_{s}$ varies from 0.979 to 0.989.  

In Table 4 we show how the Prescription II inflation predictions vary with $m_{t}$ for three $T$ vector quarks with fixed $m_{Q} = 3 \TeV$. We see that the value of $n_{s}$ is greater than 0.984 and $r$ is greater than 0.01 over the whole 2-$\sigma$ experimental range of $m_{t}$.

\subsection{${\bf n_{Q} = 2}$}

We next consider the effect of fewer VLQs. In this case the upper bounds on $m_{Q}$ are smaller, increasing the likelihood that they will be observed in future colliders. 
For two $T$ vector quarks and the mean value $m_{t} = 172.69 \GeV$, the absolute stability upper bound on $m_{Q}$ is 4.4 TeV, whilst the finite temperature stability upper bound is 5.8 TeV. The corresponding upper bounds for the case of two $B$ vector quarks are 2.6 TeV and 2.8 TeV respectively. In Table 5 and Figure 8 we show how the stability bounds vary with $m_{t}$ for the case of two $T$ vector quarks.

In Figure 9 and Table 6 we show $r$ and $n_{s}$ for $m_{Q}$ varying from 1 TeV to 5 TeV for the case of Prescription II with two $T$ vector quarks and $m_{t}$ equal to its mean value. As in the $n_{Q} = 3$ case, the observational upper bound on $r$ imposes a stronger constraint on $m_{Q}$ that stability of the potential, with $m_{Q} \lae 2.4$ TeV compared with 5.8 TeV from finite temperature stability. The values of $r$ are in the range 0.017 to 0.04 and $n_{s}$ is in the range 0.988 to 0.990 for $m_{Q}$ in the range 1 TeV to 2.4 TeV.  

In Figure 10 and Table 7 we show the corresponding results for the case of two $B$ vector quarks. In this case the observational upper bound on $r$ imposes an upper bound on $m_{Q}$ of 1.75 TeV, compared to 2.8 TeV from finite temperature stability. The values of $r$ are in the range 0.016 to 0.04 and $n_{s}$ is in the range 0.987 to 0.990 for $m_{Q}$ in the range 1 TeV to 1.75 TeV. 

\subsection{${\bf n_{Q} = 1}$}

For the case of a single vector quark, we find stability is only possible if $m_{t}$ is at its negative 2-$\sigma$ value or less. For the case of a single $T$ vector quark, the absolute stability and the finite temperature stability bounds on $m_{Q}$  are both $1.6 \TeV$. For the case of a single $B$ vector quark there is stability only at $m_{Q} \leq 1.0 \TeV$. For Prescription II we find that there is no value of $m_{Q} \geq 1 \TeV$ for which $r$ is less than the observational upper bound. Therefore $n_{Q} = 1$ is possible for Prescription I but it is ruled out for Prescription II.

\section{Conclusions} 

Higgs Inflation is a minimal approach to inflation, using only the SM fields and TeV-scale extensions to achieve inflation. Here we have considered a purely fermionic extension with the addition of isosinglet vector-like quarks. This extension can be thought of as a continuation of the structure of the SM, with a single scalar multiplet plus fermions.

It is likely that the VLQ mass is in the range 1-10 TeV.  This will be true if the $t$-quark mass is at its mean value or in the upper half of its 2-$\sigma$ range. The $m_{Q}$ upper bounds are also lower for smaller numbers of vector quarks. 
Thus there is good reason to hope that proposed future particle colliders such as the HL-LHC and FCC-hh/SppC will be able to detect VLQs if VLQ stabilised Higgs Inflation is correct \cite{bel,fc}.  

The inflation observables predicted by the model strongly depend upon the conformal frame in which the theory is renormalised.  Prescription I produces predictions for $n_{s}$ and $r$ that are essentially identical to the classical predictions of Higgs Inflation, $n_{s} = 0.966$ and $r = 3.3 \times 10^{-3}$, with the spectral index in good agreement with the conventional $\Lambda$CDM CMB value. However, in light of the Hubble tension, it is not clear that the $\Lambda$CDM value of $n_{s}$ is correct. Solutions of the tension that modify the sound horizon, in particular Early Dark Energy, typically favour values in the range 0.980-0.995. It may therefore be significant that the Prescription II prediction of $n_{s}$ is also in the range 0.980-0.995. Moreover, this is accompanied by a values of the tensor-to-scalar ratio in the range 0.01-0.04, with the upper bound on $m_{Q}$ coming from the present observational upper bound on $r$ rather than from Higgs potential stability. Therefore Prescription II Higgs Inflation with VLQs typically predicts 1-10 TeV VLQ masses, a value of $n_{s}$ compatible with early-time solutions to the Hubble tension, and primordial gravitational waves that will be easily detectable by the next generation of CMB observations.


   \renewcommand{\theequation}{A-\arabic{equation}}
 \setcounter{equation}{0}  

\section*{Appendix A: VLQ modifications of the SM RG equations} 

The 1-loop and leading 2-loop modifications to the SM RG equations due to $n_{Q}$  VLQs in the $({\bf 3}, {\bf 1}, Y_{Q})$  representation are \cite{vlqs1} 
\be{a1} \Delta \beta_{g_{3},\,1-loop} = \frac{g_{3}^{2}}{16 \pi^{2}} \left(\frac{2}{3} n_{Q} \right) ~,\ee 
\be{a2} \Delta \beta_{g_{3},\, 2-loop} = \frac{g_{3}^{5}}{\left(16 \pi^{2}\right)^{2}} \left(10 n_{Q}\right) ~,\ee 
\be{a3} \Delta \beta_{g',\, 1-loop} = \frac{g'^{\,3}}{\left(16 \pi^{2}\right)^{2}} \left(4 n_{Q} Y_{Q}^{2} \right) ~,\ee
and  
\be{a4} \Delta \beta_{y_{t},\, 2-loop} = \frac{y_{t} g_{3}^{4} }{\left(16 \pi^{2}\right)^{2}} \left(\frac{40}{9} n_{Q} \right) ~.\ee

   \renewcommand{\theequation}{B-\arabic{equation}}
 \setcounter{equation}{0}  

\section*{Appendix B: $N_{*}$ for Instant Reheating}

The reheating temperature, assuming instant reheating and an Einstein frame energy density at the end of inflation given by $V_{end} = V_{E}(\sigma_{end}) \approx \lambda_{h} M_{Pl}^{4}/(4 \xi^{2})$, where $\lambda_{h}$ and $\xi$ are calculated at $\mu = \phi_{end} \approx M_{Pl}/\sqrt{\xi}$, is given by 
\be{e18} T_{R} = \left(\frac{30 \, V_{end}}{\pi^{2} g(T_{R})}\right)^{1/4}   ~,\ee
where $g(T_{R}) \approx 106.75 + 10.5 n_{Q}$ for the SM plus $n_{Q}$ isosinglet VLQs.

The number of e-foldings $N_{*}$ at the pivot scale, $k_{*}$, is obtained from 
\be{e19} \frac{2 \pi}{k_{*}} \left( \frac{a_{N}}{a_{0}} \right) \equiv  \frac{2 \pi}{k_{*}} \left(\frac{g(T_{0})}{g(T_{R})}\right)^{1/3} \left(\frac{T_{0}}{T_{R}}\right)e^{-N} = H^{-1} ~,\ee
where $T_{0}$ is the present CMB temperature, $g(T_{i})$ are the effective number of relativistic degrees of freedom, and $k_{*} = 0.05 \, {\rm Mpc}^{-1}$ is the Planck pivot scale. During inflation 
\be{e20} H \approx \left(\frac{V_{end}}{3 M_{Pl}^{2}}\right)^{1/2}  ~.\ee 
Therefore 
 \be{e21} N_{*} = \ln \left( \frac{2 \pi T_{0}}{k_{*}} \left(\frac{g(T_{0})}{g(T_{R})}\right)^{1/3}
\left(\frac{\pi^{2} g(T_{R})}{270} \right)^{1/4} \left(\frac{V_{end}^{1/4}}{M_{Pl}} \right) \right)  ~.\ee
For $n_{Q} = 3$ we obtain 
\be{e22} N_{*} = 57.4 + \frac{1}{4} \ln \left(\frac{V_{end}}{10^{64} \GeV^{4}} \right) ~.\ee

\end{document}